\documentclass[11pt,letterpaper]{article}

\usepackage{amsmath,amsfonts,amsthm,latexsym,amssymb,color}
\usepackage{hyperref}

\newtheorem{theorem}{Theorem}[section]
\newtheorem{lemma}[theorem]{Lemma}

\newtheorem{claim}[theorem]{Claim}
\newtheorem{fact}[theorem]{Fact}

\newtheorem{definition}{Definition}
\newtheorem{defn}[definition]{Definition}
\theoremstyle{definition}

\newtheorem{rem}{Remark} 
\newtheorem{remark}[rem]{Remark}

\newcommand{\secref}[1]{Section~\ref{sec:#1}}

\newcommand{\comment}[1]{}
\newcommand{\ignore}[1]{}

\newenvironment{CompactEnumerate}{
  \begin{list}{\arabic{enumi}.}{%
      \usecounter{enumi} %
      \setlength{\itemsep}{1pt}
      }}  
  {\end{list}}

\newcommand{\ket}[1]{|#1  \rangle}
\newcommand{\bra}[1]{\langle#1 |}
\newcommand{\proj}[1]{|{#1} \rangle\langle {#1} |}

\newcommand{\set}[1]{\left\{ {#1} \right\}}
\newcommand{\paren}[1]{\left( {#1} \right)}

\newcommand{\ith}{i^{th}}

\newcommand{\tr}{\mbox{Tr}}

\newcommand{\ze}{\mathbb{Z}}
\newcommand{\ce}{\mathbb{C}}

\newcommand{\eps}{\epsilon}
\newcommand{\logeps}{\log\paren{\frac{1}{\eps}}}

\newcommand{\bit}[1]{\{0,1\}^{#1}}

\newcommand{\from}{\leftarrow}

\newcommand{\half}{\frac{1}{2}}

\newcommand{\E}{{\mathbb{E}}}

\newcommand{\bbF}{{\mathbb {F}}}

\newcommand{\I}{{\cal I}}

\newcommand{\id}{\mathbb{I}}

\newcommand{\expe}[2]{\E_{{#1}} \left[ {#2} \right]}

\newcommand{\cE}{{\mathcal{E}}}

\title{Small Pseudo-Random Families of Matrices:\\
  Derandomizing Approximate
  Quantum Encryption
}

\author{Andris Ambainis\thanks{e-mail: ambainis@ias.edu. Supported by NSF grant DMS-0111298.}\\IAS, Princeton 
\and Adam Smith\thanks{e-mail:csail.mit.edu. Supported by Microsoft 
Fellowship.}\\MIT}

\begin{document}
\maketitle

\begin{abstract}
  A \emph{quantum encryption scheme} (also called \emph{private
    quantum channel}, or \emph{state randomization protocol}) is a
  one-time pad for quantum messages. If two parties share a classical
  random string, one of them can transmit a quantum state to the other
  so that an eavesdropper gets little or no information about the
  state being transmitted.  \emph{Perfect} encryption schemes leak no
  information at all about the message.  \emph{Approximate} encryption
  schemes leak a non-zero (though small) amount of information
  but require a shorter shared random key. Approximate schemes with
  short keys have been shown to have a number of applications in
  quantum cryptography and information theory \cite{HLSW}.
  
  This paper provides the first deterministic, polynomial-time
  constructions of quantum approximate encryption schemes with short
  keys. Previous constructions \cite{HLSW} are probabilistic---that
  is, they show that if the operators used for encryption are chosen
  at random, then with high probability the resulting protocol will be
  a secure encryption scheme.  Moreover, the resulting protocol
  descriptions are exponentially long.  Our protocols use keys of the
  same length as (or better length than) the probabilistic
  constructions; to encrypt $n$ qubits approximately, one needs
  $n+o(n)$ bits of shared key~\cite{HLSW}, whereas $2n$ bits of key
  are necessary for perfect encryption~\cite{AMTW00}.
  
  An additional contribution of this paper is a connection between
  classical combinatorial derandomization and constructions of
  pseudo-random matrix families in a continuous space.


  
\end{abstract}

\section{Introduction}
\label{sec:intro}

A \emph{quantum encryption scheme} (or \emph{private quantum channel},
or \emph{state randomization protocol}) allows Alice, holding a
\emph{classical} key\footnote{Classical keys are inherently easier to
  store, distribute and manipulate, since they can be copied. More
  subtly, encryption with a shared quantum key is in many ways a dual
  problem to encryption with a classical key; see \cite{HLSW,BHLSW}
  for more discussion.}, to scramble a quantum state and send it to
Bob (via a quantum channel) so that (1) Bob, given the key, can
recover Alice's state exactly and (2) an adversary Eve who intercepts
the ciphertext learns nothing about the message, as long as she
doesn't know the key.

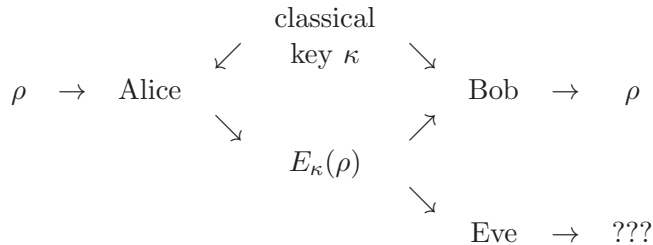
\begin{center}
  \begin{tabular}{ccccccccc}
    && &&classical  &&\\
    && & $\swarrow$ &key $\kappa$& $\searrow$ &\\
    $\rho$ & $\rightarrow$& Alice &&& & Bob &$\rightarrow$ & $\rho$ \\
    && & $\searrow$ &  &$\nearrow$\\
    && && $E_\kappa(\rho)$&&\\
    && &&&$\searrow$\\
    && &&&& Eve & $\rightarrow$ & ???
  \end{tabular}
\end{center}


There are two variants of this definition.  An encryption scheme is
called \emph{perfect} if Eve learns zero information from the
ciphertext, and \emph{approximate} if Eve can learn some non-zero
amount of information.  A perfect encryption ensures that the
distributions (density matrices) of ciphertexts corresponding to
different messages are exactly identical, while an approximate scheme
only requires that they be very close; we give formal definitions
further below.  In the classical case, both perfect and approximate
encryption require keys of roughly the same length---$n$ bits of key
for $n$ bits of message. In the quantum case, the situation is
different.

For perfect encryption, Ambainis et al.  \cite{AMTW00} showed that
$2n$ bits of key are necessary and sufficient to encrypt $n$ qubits.
The construction consists of applying two classical one-time
pads---one in the ``standard'' basis $\set{\ket{0},\ket{1}}$ and
another in the ``diagonal'' basis $\{\frac{1}{\sqrt{2}} (\ket{0} +
\ket 1) , \frac{1}{\sqrt{2}} (\ket{0} - \ket 1) \}$.


Approximate encryption was studied by Hayden et al. \cite{HLSW}. They
introduced an additional, useful relaxation: they show that if the
plaintext is not entangled with Eve's system to begin with, then one
can get \emph{approximate} quantum encryption using only $n+o(n)$ bits
of key---roughly half as many as are necessary for perfect encryption.
\footnote{The result of \cite{HLSW} highlights an error in the proof
  of a lower bound on key length of authentication schemes in
  \cite{BCGST02}. The results of that paper remain essentially
  correct, but the definition of authentication requires some
  strengthening, and the proof of the
  lower bound is more involved.}  %
The assumption that Eve's system is unentangled with the message is
necessary for this result; othwerwise roughly $2n$ bits are needed,
even for approximate encryption.  The assumption holds in the quantum
counterpart of the one-time pad situation (one party prepares a
quantum message and sends it to the second party, using the encryption
scheme) as long as the message is not part of a larger cryptographic
protocol. The relaxation also has a host of less cryptographic
applications, for example: constructing efficient quantum data hiding
schemes in the LOCC (local operation and classical communication)
model; exhibiting ``locked'' classical correlations in quantum states
\cite{HLSW}; relaxed authentication of quantum states using few bits
of key \cite{leung}; and transmitting quantum states over a classical
channel using $n+o(n)$ bits of communication, rather than the usual
$2n$ bits required for quantum teleportation \cite{BHLSW}.

The previous constructions of approximate encryption schemes with a
shorter key are probabilistic. Specifically, Hayden et al. \cite{HLSW}
showed that a \emph{random} set of $2^{n+o(n)}$ unitary matrices leads
to a good encryption scheme with high probability (to encrypt, Alice
uses the key to choose one of the matrices from the set and applies
the corresponding operator to her input).  However, verifying that a
particular set of matrices yields a good encryption scheme is not
efficient; even writing down the list of matrices is prohibitive,
since there are exponentially many of them.

This paper presents the first polynomial time constructions of
approximate quantum encryption schemes (to relish the oxymoron:
derandomized randomization protocols). The constructions run in
time $O(n^2)$ when the message $\rho$ consists of $n$ qubits.
That is, given the key and the input message, Alice can produce the
output using $O(n^2)$ steps on a quantum computer. The key length we
achieve is slightly better than that of the probabilistic construction
of \cite{HLSW}.  Our results apply to the trace norm on matrices;
exact results are stated further below.

The main tools in our construction are small-bias sets \cite{NN} of
strings in $\bit{2n}$. Such sets have proved useful in derandomizing
algorithms, constructing short PCPs \cite{BSVW03} and the encryption
of high-entropy messages \cite{RW02}.  Thus, one of the contributions
of this paper is a connection between classical combinatorial
derandomization and constructions of pseudo-random matrix families in
a continuous space. Specifically, we connect Fourier analysis over
$\ce^{\ze_2^{2n}}$ to Fourier analysis over the matrices $\ce^{2^n\times
  2^n}$. This parallels, to some extent, the connection between
quantum error-correcting codes over $n$ qubits and classical codes
over $GF(4)^n$.

\subsection{Definitions and Previous Work}
\label{sec:defs}

We assume that the reader is familiar with the basic notation of
quantum computing (see \cite{NC} for an introduction). Syntactically,
an approximate quantum encryption scheme is a set of $2^k$ invertible
operators $\set{E_\kappa | \kappa \in \bit{k}}$.  The $E_\kappa$'s may
be unitary, but need not be: it is sufficient that one be able to
recover the input $\rho$ from the output $E_\kappa(\rho)$, which may
live in a larger-dimensional space than $\rho$. Each $E_\kappa$ takes
$n$ qubits as input and produces $n'\geq n$ qubits of output. If
$n'=n$ then each operator $E_\kappa$ corresponds to a unitary matrix:
$E_\kappa(\rho) = U_\kappa \rho U_\kappa^\dagger$.

For an input density matrix\footnote{Recall that for a pure state
  $\ket\phi$, the density matrix $\rho$ is $ \ket\phi\bra\phi$.}
$\rho$, the density matrix of the ciphertext from the adversary's
point of view is:
$$\cE(\rho)= \expe{\kappa}{E_\kappa(\rho)} = \frac{1}{2^k}\sum_{\kappa
  \in \bit{k}} E_\kappa (\rho)$$

When the scheme is length-preserving, this yields
$$\cE(\rho) = \frac{1}{2^k} \sum_{\kappa} U_\kappa\rho U_\kappa^\dagger$$

\begin{defn}
  The set of operators $\set{E_\kappa}$ is an approximate quantum
  encryption scheme (state randomization scheme) with error $\eps$ on
  $n$ qubits if
  \begin{equation} \label{eq:def}
    \mbox{for all density matrices $\rho$ on $n$ qubits:}
    \qquad D(\cE(\rho), \frac{1}{2^{n'}}I) = \left\|\cE(\rho) -
    \frac{1}{2^{n'}} I \right \|_tr \leq \eps.
  \end{equation}
  
\end{defn}

Here $D(\cdot,\cdot)$ refers to the trace distance between density
matrices.  The trace norm of a matrix $\sigma$ is the trace of the
absolute value of $\sigma$ (equivalently, the sum of the absolute
values of the eigenvalues).  The \emph{trace distance} between two
matrices $\rho,\sigma$ is
$$D(\rho,\sigma) = \|\rho-\sigma\|_{tr} = \tr(| \rho-\sigma| )$$

This norm plays the same role for quantum states as statistical
difference does for probability distributions: the maximum probability
of distinguishing between two quantum states $\rho,\sigma$ via a single
measurement is
$$\textstyle \half + \frac{1}{4}D(\rho,\sigma).$$

Hayden et al. \cite{HLSW} actually considered randomization
schemes with respect to two norms: the $\infty$-norm (``operator
norm'') and the trace norm. In this paper, we consider schemes for the
trace norm, though our proofs go through the Frobenius norm.
Constructing explicit randomization schemes for the $\infty$-norm
remains an interesting open problem.

\begin{remark}
  This definition of quantum encryption implicitly assumes that the
  message state $\rho$ is not entangled with the adversary's system.
  Without that assumption the definition above is not sufficient, and
  it is \emph{not} possible to get secure quantum encryption using
  $n(1+o(1))$ bits of key (roughly $2n$ bits are provably necessary).
  Thus, this sort of construction is not universally useful in
  cryptographic contexts, but nevertheless has many applications
  (described above).
\end{remark}

\paragraph{Previous Work}

Ambainis et al. \cite{AMTW00} considered perfect encryption; this
corresponds to the case where $\eps=0$. The choice of matrix norm is
irrelevant there, since $\cE(\rho)=\frac{1}{2^{n'}}I$. As mentioned
above, they showed that $2n$ bits of key are necessary and sufficient.
The construction uses the key to choose one of $2^{2n}$ Pauli
operators (defined below) and applies that to the input state.

Hayden et al. \cite{HLSW} showed that a set of $O(n^2 2^n /
\eps^2)$ unitary operators suffices (for both the $\infty$-norm and
the trace norm). For the trace norm, they even showed that a random
set of Pauli matrices (see below) would suffice. This means that for
encrypting $n$ qubits, they presented a non-polynomial-time scheme
requiring $n + 2\log n + 2\logeps +O(1)$ bits of key.

\subsection{Our Results}

We present three explicit, polynomial time constructions of
approximate state randomization protocols for the trace norm. All are
based on exisiting constructions of $\delta$-biased sets
\cite{NN,AGHP,ABNNR}, or on families of sets with small average bias.
The three constructions are explained and proven secure in Sections
\ref{sec:smallbias}, \ref{sec:shorter} and \ref{sec:hybrid}, resepctively.

The first construction is length-preserving, and requires 
$$
n + 2\log n + 2\logeps + O(1)$$
bits of key, thus matching the
performance of the non-explicit construction. The second construction
is length-doubling: it encodes $n$ qubits into $n$ qubits and $2n$
classical bits but uses a shorter key: only 
$$
n + 2\logeps$$
bits of key are required. Both of these constructions
are quite simple, and are proven secure using the same
Fourier-analytic technique.

The final construction has a more sophisticated proof, but allows for
a length-preserving scheme with slightly better dependence on the
number of qubits: 
$$
n+ \min\set{2\log n +2\logeps , \log n + 3\logeps } + O(1)$$
bits of
key. The right-hand term provides a better bound when $\eps >
\frac{1}{n}$.

More generally, Fourier analysis over the cube $\bit{n}$ has provided
a rich set of tools for understanding classical boolean functions and
distributions on $\bit{n}$. We hope the ideas in this paper indicate
how some of the classical results can be transposed to yield new
results in quantum information theory.

\section{Preliminaries}
\label{sec:prelims}

\paragraph{Small-Bias Spaces}

The bias of a random variable $A$ in $\bit{n}$ with respect to a
string $\alpha\in\bit{n}$ is the distance from uniform of the bit
$\alpha\odot A$, where $\odot$ refers to the standard dot product on $\ze_2^n$:
$$\hat A (\alpha)= \expe{A}{(-1)^{\alpha\odot A}} = 2\Pr[\alpha\odot A = 0]
- 1.$$
The
function $\hat A$ is the Fourier transform of the probability mass function of
the distribution, taken over the group $\ze_2^n$.

The bias of a set $S\in\bit{n}$ with respect to $\alpha$ is simply the bias of
the uniform distribution over that set. A set $S$ is called $\delta$-biased if
the absolute value of its bias is at most $\delta$ for all $\alpha\neq 0^n$.

Small-bias sets were first considered in derandomization theory by
Naor and Naor \cite{NN}. Alon, Bruck et al. (ABNNR, \cite{ABNNR}) gave
explicit (i.e.~deterministic, polynomial-time) constructions of
$\delta$-biased sets in $\bit{n}$ with size $O(n/\delta^3)$.
Constructions with size $O(n^2/\delta^2)$ were provided by Alon,
Goldreich, et al. (AGHP, \cite{AGHP}). The AGHP construction is better
when $\delta = o(1/n)$.  In both cases, the $\ith$ string in a set can
be constructed in roughly $n^2$ time (regardless of $\delta$).

One can sample a random point from a $\delta$-biased space over $\bit{n}$ using
either $\log n + 3\log (1/\delta)+O(1)$ bits of randomness (using ABNNR) or using
$2\log n + 2\log(1/\delta)$ bits (using AGHP).  

\paragraph{Small-bias Set Families}

One can generalize small bias to \emph{families} of sets (or random
variables) by requiring that on average, the bias of a random set from
the family with respect to every $\alpha$ is low (Dodis and Smith
\cite{DS03}).  Specifically, the expectation of the \emph{squared}
bias must at most $\delta^2$.  Many results on $\delta$-biased sets
also hold for $\delta$-biased families, which are easier to construct.

\begin{defn}
  A \emph{family} of random variables (or sets) $\set{A_i}_{i\in I}$ is
  $\delta$-biased if 
  
  \begin{center}
    $\sqrt{\expe{i\from I}{\hat A_i(\alpha)^2}} \leq \delta $ for all $ \alpha
    \neq 0^n.$
\end{center}
\end{defn} 

Note that this is \emph{not} equivalent, in general, to requiring that
the expected bias be less than $\delta$.  There are two important
special cases:
\begin{CompactEnumerate}
\item If $S$ is a $\delta$-biased set, then $\set{S}$ is a
  $\delta$-biased set family with a single member;

\item A family of linear spaces $\set{C_i}_{i\in I}$ is {\em
    $\delta$-biased} if no particular word is contained in the dual
  $C_i^\perp$ of a random space $C_i$ from the family with high
  probability.  Specifically:
  $$\hat C_i(\alpha) = \left\{
    \begin{array}{lcl} 
      0 &\mathrm{if} & \alpha\not\in C_i^\perp \\
      1 &\mathrm{if} & \alpha\in C_i^\perp 
    \end{array}\right . $$
  Hence a family of codes is $\delta$-biased if and only if
  $\Pr_{i\from I}[\alpha\in C_i^\perp] \leq \delta^2$, for every
  $\alpha\neq 0^n$.  Note that to meet the definition, for linear
  codes the expected bias must be at most $\delta^2$, while for a
  single set the bias need only be $\delta$.
\end{CompactEnumerate}

One can get a good $\delta$-biased family simply by taking $\set{C_i}$
to be the set of all linear spaces of dimension $k$. The probability
that any fixed non-zero vector $\alpha$ lies in the dual of a random
space is exactly $\delta^2 = \frac{2^{n-k}-1}{2^n-1}$, which is at
most $2^{-k}$.

One can save some randomness in the choice of the space using a
standard pairwise independence construction. View $\bit{n}$ as
$GF(2^n)$, and let $K \subseteq GF(2^n)$ be an additive subgroup of size
$2^k$. For every non-zero string $a$, let the space $C_a$ be given by
all multiples $a\kappa$, where $\kappa\in K$. The family $\set{C_a\ |\ 
  a \in GF(2^n), a\neq 0}$ has the same bias as the set of all linear
spaces ($\delta<2^{-k/2}$), and $n$ bits of randomness are needed to
choose a set in the family. 

\paragraph{Entropy of Quantum States}

As with classical distributions, there are several ways to measure the
entropy of a quantum density matrix. We'll use the analogue of
collision entropy (a.k.a. Renyi entropy).

For a classical random variable $A$ on $\bit{n}$, the collision
probability of two independent samples of $X$ is $p_c=\sum_a
\Pr[A=a]^2$. The Renyi entropy of $A$ is $H_2(A)=-\log p_c$.

For a quantum density matrix $\rho$, the analogous quantity is
$H_2(\rho) = -\log \tr(\rho^2)$. If the eigenvalues of $\rho$ are
$\set{p_x}$, then the eigenvalues of $\rho^2$ are $\set{p_x^2}$, and
so $\tr(\rho^2)$ is exactly the collision probability of the
distribution obtained by measuring $\rho$ in a basis of eigenvectors.

\begin{fact}
\label{fact:trace2}
  If $\rho$ describes a state in $d$-dimensional space
   and $\tr(\rho^2)\leq \frac{1}{d}
  (1+\eps^2)$, then $D(\rho, \frac{1}{d}\id) \leq \eps$.
\end{fact}


\paragraph{Pauli matrices}

The $2\times 2$ Pauli matrices are generated by the matrices:

$$X = \left(
\begin{array}{cc}
0&1 \\ 1&0
\end{array}
\right)
\qquad
Z =  \left(
\begin{array}{cc}
1&0 \\ 0&-1
\end{array}
\right)
$$

The Pauli matrices are the four matrices $\set{\id,X,Z,XZ}$. These form
a basis for the space of all $2\times 2$ complex matrices. Since
$XZ=-ZX$, and $Z^2=X^2=1$, the set generated by $X$ and $Z$ is given
by the Pauli matrices and their opposites: $\set{\pm \id, \pm X,\pm
  Z,\pm XZ}$.

If $u$ and $v$ are $n$-bit strings, we denote the corresponding tensor
product of Pauli matrices by $X^uZ^v$. That is, if we write
$u=(u1,...,u_n)$ and $v=(v_1,...,v_n)$, then 
$$X^{u}Z^{v} = X^{u_1}Z^{v_1}\otimes \cdots \otimes
X^{u_n}Z^{v_n}.$$

(The strings $x$ and $z$ indicate in which positions of the tensor
product $X$ and $Z$ appear, respectively.)  The set $\set{X_uZ_v \ |\ 
  u,v \in \bit{n}}$ forms a basis for the $2^n\times 2^n$ complex
matrices.  The main facts we will need are given below:

\begin{enumerate}
\item Products of Pauli matrices  obey the group structure of
  $\bit{2n}$ up to a minus sign. That is, $(X^uZ^v)(X^aZ^b) = (-1)^{a\odot v}X^{u\oplus a}Z^{v\oplus
    b}$.
\item Any pair of Pauli matrices either commutes or anti-commutes. That is,\\
  $(X^uZ^v)(X^aZ^b) = (-1)^{u\odot b + v\odot a} (X^aZ^b)(X^uZ^v)$.
  
\item The trace of $X^uZ^v$ is 0 if $(u,v)\neq 0^{2n}$ (and
  otherwise it is $\tr(\id)=2^n$).
\item $(X^uZ^v)^\dagger = Z^vX^u = (-1)^{u\odot v} X^uZ^v$
\end{enumerate}

\paragraph{Pauli matrices and Fourier Analysis}

\newcommand{\twon}{{\textstyle \frac{1}{2^n}}}

The Pauli matrices form a basis for the set of all $2^n\times2^n$
matrices. Given a density matrix $\rho$, we can write
$$\rho = \sum_{u,v\in\bit{n}} \alpha_{u,v}X^uZ^v.$$

This basis is orthonormal with respect to the inner product given by
$\twon\tr(A^\dagger B)$, where $A,B$ are square matrices. That is:
$$\twon\tr((X^uZ^v)^\dagger X^aZ^b) = \delta_{a,u}\delta_{b,v}.$$

Thus, the usual arithmetic of orthogonal bases (and Fourier analysis)
applies.  One can immediately deduce certain properties of the
coefficients $\alpha_{u,v}$ in the decomposition of a matrix $\rho$.
First, we have a formula for $\alpha_{u,v}$:
$$\alpha_{u,v} = \twon\tr(Z^vX^u \rho).$$

Second, the squared norm of $\rho$ is given by the squared norm of the
coefficients. 
$$\twon\tr(\rho^\dagger\rho)= \sum_{u,v}|\alpha_{u,v}|^2$$

Since $\rho$ is a density matrix, it is Hermitian
($\rho^\dagger=\rho$). One can use this fact, and our formula for the
coefficients $\alpha_{u,v}$, to get a compact formula for the entropy
in terms of the decomposition in the Pauli basis:
$$\tr(\rho ^2 )  =  \frac{1}{2^{n}}\sum_{u,v} |\tr(X^uZ^v\rho)|^2.$$

\section{State Randomization and Approximate Encryption}

\subsection{Encrypting with a Small-Bias Space}
\label{sec:smallbias}

The ideal quantum one-time pad applies a random Pauli matrix to the
input \cite{AMTW00}. Consider instead a scheme which first chooses a
$2n$-bit string from some set with small bias $\delta$ (we will set
$\delta$ later to be roughly $\delta=\epsilon2^{-n/2}$). If the set of
strings is $B$ we have:
$$\cE(\rho_0) = \frac{1}{|B|}\sum_{(a,b)\in B} X^aZ^b\rho_0 Z^bX^a =
\expe{a,b}{ X^aZ^b\rho_0 Z^bX^a}$$
That is, we choose the key from the
set $B$, which consists of $2n$-bit strings. To encrypt, we view a
$2n$-bit string as the concatenation $(a,b)$ of two strings of $n$
bits, and apply the corresponding Pauli matrix.

(The intuition comes from the proof that Cayley graphs based on
$\eps$-biased spaces are good expanders: applying a Pauli operator
chosen from a $\delta$-biased family of strings to $\rho_0$ will cause
all the Fourier coefficients of $\rho_0$ to be reduced by a factor of
$\delta$, which implies that the collision entropy of $\rho_0$ also
gets multiplied by $\delta$. We expand on this intuition below. )

As a first step, we can try to see if a measurement given by a Pauli
matrix $X^uZ^v$ can distinguish the resulting ciphertext from a totally
mixed state. More explicitly, we perform a measurement which projects
the ciphertext onto one of the two eigenspaces of the matrix $X^uZ^v$.
We output the corresponding eigenvalue. (All Pauli matrices have two
eigenvalues with eigenspaces of equal dimension.  The eigenvalues are
always either $-1$ and $1$ or $-i$ and $i$.)

To see how well a particular Pauli matrix $X^uZ^v$ will do at
distinguishing, it is sufficient to compute
$$|\tr(X^uZ^v\cE(\rho_0))|.$$

This is exactly the statistical difference between the Pauli
measurement's outcome and a uniform random choice from the two
eigenvalues. We can compute $\tr(X^uZ^v\cE(\rho_0))$ explicitly:

\begin{eqnarray*}
  \tr(X^uZ^v\cE(\rho_0)) &=& \tr\paren{X^uZ^v\expe{(a,b)\in
    B}{X^aZ^b\rho_0 Z^vX^u}}
\\
& =&
\expe{a,b}{\tr(X^uZ^vX^aZ^b\rho_0Z^bX^a)} 
\\
&=&
\expe{a,b}{\tr(Z^bX^aX^uZ^vX^aZ^b\rho_0)}
\\
& =&\expe{a,b}{(-1)^{a\odot v + b
    \odot u}}\tr(X^uZ^v\rho_0)
\end{eqnarray*}

Since $a\odot v + b \odot u$ is linear in the concatenated $2n$-bit
vector $(a,b)$, we can take advantage of the small bias of the set $B$
to get a bound:

$$|\tr(X^uZ^v\cE(\rho_0))| \leq \delta|\tr(X^uZ^v\rho_0)|$$

Equivalently: if we express $\rho_0$ in the basis of matrices
$X^uZ^v$, then each coefficient shrinks by a factor of at least
$\delta$ after encryption. We can now bound the distance from the
identity by computing $\tr(\cE(\rho_0)^2)$:

$$\tr(\cE(\rho_0)^2) = \frac{1}{2^n}\sum_{u,v}
|\tr(X^uZ^v\cE(\rho_0))|^2 \leq \frac{1}{2^n} +
\frac{\delta^2}{2^n}\sum_{(u,v)\neq 0^{2n}}|\tr(X^uZ^v\rho_0)|^2 \leq
\frac{1}{2^n}(1+\delta^22^n \tr(\rho_0^2) )$$

Setting $\delta = \sqrt{2}\epsilon 2^{-n/2}$, we get approximate
encryption for all states (since $\tr(\rho_0^2)\leq 1$).  Using the
constructions of AGHP \cite{AGHP} for small-bias spaces, we get a
polynomial-time scheme that uses $n + 2\log n + 2\logeps$ bits of key.

\subsection{A  Scheme with Shorter Key Length}
\label{sec:shorter}

We can improve the key length of the previous scheme using
$\delta$-biased \emph{families} of sets. The tradeoff is that the
resulting states are longer: the ciphertext consists of $n$ qubits and
$2n$ classical bits.
In classical terms, the encryption algorithm uses additional
randomness which is not part of the shared key; in the quantum
computing model, however, that randomness is ``free'' if one is
allowed to discard ancilla qubits. 

\begin{lemma}
  If $\set{A_i}_{i\in\I}$ is a family of subsets of $\bit{2n}$ with
  average square bias $\delta^2$, then the operator
  $$\cE(\rho_0) = \expe{i\in \I}{\proj{i}\otimes \expe{ab \in A_i}
    {X^aZ^b \rho_0 Z^bX^a}}$$
  is an approximate encryption scheme for
  $n$ qubits with leakage $\eps$ whenever $\delta \leq \eps 2^{-n/2}$.
\end{lemma}

Before proving the lemma, we give an example using the small-bias set
family from the preliminaries. View the key set $\bit{k}$ as an
additive subgroup $K$ of the field $\bbF=GF(2^{2n})$.  For every
element $a\in\bbF$, define the set $C_a =\set{a\kappa | \kappa\in K}$.
The family ${C_a}$ has bias $\delta<2^{-k/2}$. The corresponding
encryption scheme takes a key $\kappa\in\bit{k}\subseteq GF(2^{2n})$:

$$\cE (\rho_0;\kappa) = \left[
\begin{array}{l}
\mbox{Choose}\ \alpha \from_R GF(2^{2n})\setminus \set{0} \\
\mbox{Write}\ \alpha\kappa = (a,b),\ \mbox{where}\ a,b \in\bit{n}\\
\mbox{Output the classical string}\ \alpha\ \mbox{and the quantum state}\  X^aZ^b\rho_0 Z^b X^a
\end{array}\right.$$

With a quantum computer, random bits are not really necessary for
choosing $\alpha$; it is sufficient to prepare $2n$ EPR pairs and
discard one qubit from each pair.  For the scheme to be secure, the
bias $\delta$ should be at most $\sqrt{\eps/2^{n}}$, and so the key
only needs to be $n+2\logeps$ bits long. The main disadvantage is that
the length of the ciphertext has increased by $2n$ classical bits.


\begin{proof}
  As before, the proof will use elementary Fourier analysis over the
  hypercube $\ze_2^{2n}$, and intuition comes from the proof that
  Cayley graphs based on $\eps$-biased set families are also
  expanders.
  
  Think of the output of the encryption scheme as a single quantum
  state consisting of two systems: the first system is a classical
  string describing which member of the $\delta$-biased family will be
  used. The second system is the encrypted quantum state. To complete
  the proof, it is enough to bound the collision entropy of the entire
  system by $\frac{1}{2^n{|\I|}}(1+2\eps^2)$.
  
  For each $i\in\I$ (that is, for each member of the set family), let
  $\rho_i$ denote the encryption of $\rho_0$ with a random operator
  from the set $A_i$.  The first step of the proof is to show that the
  collision entropy of the entire system is equal to the average
  collision entropy of the states $\rho_i$.  

  \begin{claim}
    $\displaystyle \tr(\cE(\rho_0)^2) = \frac{1}{|\I|}\expe{i\from I}{\tr(\rho_i^2)}$
  \end{claim}

  \begin{proof}
    We can write $\cE(\rho_0) = \frac{1}{|\I|}\sum_i \ket i \bra i
    \otimes \rho_i$. Then we have
    $$\textstyle
    \tr(\cE(\rho_0)^2 ) = \frac{1}{|\I|^2} \sum_{i,j}\tr\big(
    (\ket{i}\bra i \ket j \bra j ) \otimes \rho_i \rho_j \big)$$
    Since
    $\bra i \ket j = \delta_{i,j}$, we get $\tr(\cE(\rho_0)^2 ) =
    \frac{1}{|\I|^2} \sum_i \tr(\rho_i^2)$, as desired.
  \end{proof}

  Take  any string $w=(u,v)\in \bit{2n}$, where $u,v\in\bit{n}$.  Recall
  that $\hat A_i(u,v)$ is the ordinary Fourier coefficient (over
  $\ze_2^{2n}$) of the uniform distribution on $A_i$, that is $\hat
  A_i(u,v) = \expe{a \from A_i }{(-1)^{a\odot w}}$.  From the previous
  proof, we know that
  $$\tr(X^u Z^v \rho_i)=\hat A_i(v,u)\cdot \tr(X^uZ^v \rho_0).$$
  
  We can compute the now average collision entropy of the states $\rho_i$. Using
  linearity of expectations:
\begin{eqnarray*}
   \expe{i}{\tr(\rho_i^2)} 
  &=& \expe{i}{ \twon + \twon \sum_{(u,v)\neq 0} |\tr(X^uZ^v\rho_i)|^2} 
\\
  &=& \twon+\twon \sum_{(u,v)\neq 0} \expe{i}{|\tr(X^uZ^v\rho_i)|^2}
\\
  &=&  \twon+\twon \sum_{(u,v)\neq 0} \expe{i}{\hat A_i(v,u)^2}
  |\tr(X^uZ^v\rho_0)|^2 
  \end{eqnarray*}
  The expression $ \expe{i}{\hat A_i(v,u)^2}$ is exactly the quantity
  bounded by the (squared) bias $\delta^2$. As in the previous proof,
  the entropy $\tr(\cE(\rho_0)^2)$ is bounded by $\frac{1}{2^n|\I|}(1+ \delta^22^n\tr(\rho_0^2))$. By our choice of $\delta$,
  the entropy is at most $\frac{1}{2^n|\I|}(1+\eps^2)$, and so
  $\cE(\rho_0^2)$ is within trace distance $\eps$ of the completely
  mixed state.
\end{proof}

\subsection{Hybrid Construction}
\label{sec:hybrid}

Let $d$ be a prime between $2^n$ and $2^{n+1}$.  Then, it suffices to
show how to randomize a state in a $d$-dimensional space ${\cal H}_d$
spanned by $\ket{i}$, $i\in\{0, 1, \ldots, d-1\}$, since a state on
$n$ qubits can be embedded into ${\cal H}_d$.  We define $X$ and $Z$
on this space by $X\ket{j}=\ket{(j+1)\bmod d}$ and $Z\ket{j}=e^{2\pi i
  j/d} \ket{j}$.  Notice that $X^j Z^k = e^{2\pi i (j k)/d} Z^k X^j$
and $(X^j Z^k)^{\dagger}=Z^{-k} X^{-j}$. (The definitions of $X$ and
$Z$ are different than in the previous sections, since we are
operating on a space of prime dimension).

We start with a construction that uses $n+1$ bits of randomness and 
achieves approximate encryption for $\epsilon=1$.
(Notice that this is a non-trivial security guarantee.
The trace distance between perfectly distinguishable states is 2.
Distance 1 means that the state cannot be distinguished from $\frac{I}{d}$
with success probability more than 3/4.)
We will then extend it to any $\epsilon>0$, using 
more randomness.

Let 
\[ \cE (\rho) = \frac{1}{d}\sum_{a=1}^{d-1} X^a Z^{a^2} \rho Z^{-a^2} X^{-a} .\]

\begin{claim}
\label{claim:a}
\[ Tr(\cE(\rho)^2) \leq \frac{1}{d} (1+Tr(\rho^2)) .\]
\end{claim}
 
\begin{proof}
Let $\rho'=\cE(\rho)$.
\[ Tr (\rho')^2=\sum_{ij} \rho'_{ij} (\rho'_{ij})^* =\sum_i 
\rho'_{ii} (\rho'_{ii})^* + 
\sum_{i,j:i\neq j} \rho'_{ij} (\rho'_{ij})^* .\]
The first sum is equal to $d \frac{1}{d^2}=\frac{1}{d}$ because 
$\rho'_{ii}=\frac{1}{d} \sum_{k=1}^d \rho_{kk}=\frac{1}{d}$. 
To calculate the second sum, we split it into sums
$S_t=\sum_i \rho'_{i, i+t} (\rho'_{i, i+t})^*$ 
for $t=1, 2, \ldots, d-1$. (In the indices for $\rho_{ij}$
and $\rho'_{ij}$, we use
$i+t$ as a shortcut for $(i+t)\bmod d$.) We have 
\[ \rho'_{i, i+t}=\frac{1}{d} \sum_{a=0}^{d-1} w^{a^2 t}
\rho_{i-a, i-a+t} ,\]
where $w$ is the $d^{\rm th}$ root of unity.
\[ \rho'_{i, i+t} (\rho'_{i, i+t})^* = \frac{1}{d^2} 
\left( \sum_{a=0}^{d-1} |\rho_{i+a, i+t+a}|^2 + \sum_{a, b, a\neq b}
w^{(b^2-a^2) t} \rho_{i-a, i+t-a} (\rho_{i-b, i+t-b})^* \right).\]
Therefore,
\[ S_t = \frac{1}{d} \sum_{i=1}^d |\rho_{i, i+t}|^2 +
\frac{1}{d^2} \sum_{i \neq j} c_{i, j} \rho_{i, i+t} 
(\rho_{j, j+t})^* \]
where
\[ c_{i, j}=\sum_a w^{((i+a)^2-(j+a)^2)t}=\sum_a w^{(i^2-j^2+2a(i-j))t}=
w^{(i^2-j^2)t} \sum_a w^{a*2(i-j)t} .\]
Since $d$ is a prime, $2(i-j)t$ is not divisible by $d$.
Therefore, $\sum_a w^{a*2(i-j)t}=0$, $c_{ij}=0$, 
$S_t=\frac{1}{d} \sum_{i=1}^d |\rho_{i, i+t}|^2$ and
\[ Tr((\rho')^2) = \frac{1}{d} + \frac{1}{d} \sum_{i\neq j} 
|\rho_{ij}|^2 .\]
\end{proof}

By fact \ref{fact:trace2}, $D(E(\rho), \frac{I}{d})\leq 1$.
 
We now improve this construction to any $\epsilon$.  Let $B$ be an
$\epsilon$-biased set on $m=\lceil\log d \rceil$ bits.  For $b\in \{0,
1\}^m$, define a unitary transformation $U_b$ as follows. Identify
numbers $0, 1, \ldots, d-1$ with strings $x\in\{0, 1\}^m$.  Define
$U_b\ket{x}=(-1)^{b\odot x} \ket{x}$, with $b\odot x$ being the usual
(bitwise) inner product of $b$ and $x$.  (Note that $U_b$ is just to
the $Z$ operator over a different group. It is the same $Z$ operator
used in the previous sections). Let
\[ \cE'(\rho)=\sum_{b\in B} 
U_b \rho U_b \mbox{~and~} \cE''(\rho)=\cE(\cE'(\rho)) .\]
We claim that $\cE''$ is $\epsilon$-approximate encryption scheme.
W.l.o.g., assume that $\rho$ is a pure state 
$\ket{\psi}=\sum_i c_i \ket{i}$.
Then $\rho_{ij}=c_i c^*_j$.
Let $\rho'=\frac{1}{|B|}\sum_{b\in B} U_b \rho U^{\dagger}_b$
be the result of encrypting $\rho$ by $\cE'$.
Then, 
\[ \rho'_{xy} = \frac{1}{|B|} \sum_{b\in B} (-1)^{b\odot x+ b\odot y} \rho_{xy} =
\frac{1}{|B|} \sum_{b\in B} (-1)^{b\odot ( x+y)} \rho_{xy}.\]
Since $B$ is $\epsilon$-biased,
$|\rho'_{xy}|\leq \epsilon |\rho_{xy}|$ for any $x, y$, $x\neq y$.
Therefore, $\sum_{x\neq y}|\rho'_{xy}|\leq \epsilon \sum_{x\neq y}|\rho_{xy}|$.
Together with Claim \ref{claim:a} and fact \ref{fact:trace2}, this
implies that $\cE''$ is $\epsilon$-randomizing.  The number of key
bits used by $\cE''$ is $n+\log |B|+O(1)$ which is $n+2\log n+2\log
\frac{1}{\epsilon}$ if AGHP scheme is used and $n+\log n+3\log
\frac{1}{\epsilon}$ if ABNNR is used.  The first bound is the same as
the one achieved by using small-bias spaces directly
(\secref{smallbias}).  The second bound gives a better result (as long
as $\epsilon>\frac{1}{n}$).

\section*{Acknowledgements}

We are grateful for helpful discussions with Claude Cr{\'e}peau,
Daniel Gottesman, Patrick Hayden, Debbie Leung, Sofya Raskhodnikova
and Alex Samorodnitsky.

\end{document}